# Structure evolution path of ferroelectric hafnium zirconium oxide nanocrystals under *in-situ* biasing

Yunzhe Zheng[1†], Heng Yu[2†], Tianjiao Xin[1,3†], Kan-Hao Xue[2]*, Yilin Xu[1], Zhaomeng Gao[1], Cheng Liu[1], Qiwendong Zhao[1], Yonghui Zheng[1], Xiangshui Miao[2], Yan Cheng[1,3]*

[1]Key Laboratory of Polar Materials and Devices (MOE), Department of Electronics, East China Normal University, Shanghai 200241, China.

[2]School of Integrated Circuits, Huazhong University of Science and Technology, Wuhan 430074, China.

[3]State Key Laboratory of Materials for Integrated Circuits, Shanghai Institute of Microsystem and Information Technology, Chinese Academy of Sciences, Shanghai 200050, China.

[†]Yunzhe Zheng, Heng Yu, and Tianjiao Xin contributed equally.

*Corresponding author: ycheng@ee.ecnu.edu.cn; xkh@hust.edu.cn.

**Abstract:**

Fluorite-type $HfO_2$-based ferroelectric (FE) oxides have rekindled interest in FE memories due to their compatibility with silicon processing and potential for high-density integration. The polarization characteristics of FE devices are governed by the dynamics of metastable domain structure evolution. Insightful design of FE devices for encoding and storage necessitates a comprehensive understanding of the internal structural evolution. Here, we demonstrate the evolution of domain structures through a transient polar orthorhombic (O)-*Pmn*21-like configuration via *in-situ* biasing on $TiN/Hf_{0.5}Zr_{0.5}O_2/TiN$ capacitors within spherical aberration-corrected transmission electron microscope, combined with theoretical calculations. Furthermore, it is directly evidenced that the non-FE O-*Pbca* transforms into the FE O-*Pca*2$_1$ phase under electric field, with the polar axis of the FE-phase aligning towards the bias direction through ferroelastic transformation, thereby enhancing FE polarization. As cycling progresses further, however, the polar axis collapses, leading to FE degradation. These novel insights into the intricate structural evolution path under electrical field cycling facilitate optimization and design strategies for $HfO_2$-based FE memory devices.

**Main text:**

The escalating demand for advanced memory technologies characterized by enhanced capacity and performance has catalyzed the development of innovative solutions[1]. Among these, ferroelectric (FE) memory stands out for its non-volatility, high speed, and low power consumption[2]. However, as device dimensions continue to shrink, new challenges have arisen, including scaling effects, the necessity for high-precision processing, and novel atomic-scale mechanisms[3,4]. Recently, atomic layer deposition (ALD) of $HfO_2$-based FE thin films has attracted considerable attention owing to their superior properties, potential for miniaturization, and compatibility with the complementary metal oxide semiconductor (CMOS) process, thus rejuvenating the field of FE memories[5,6].

$HfO_2$ appears in various phases, such as tetragonal (T: $P4_2/nmc$), non-FE orthorhombic ($O_I$: *Pbca* or $O_{II}$: *Pnma*), FE orthorhombic ($O_{III}$: $Pbc2_1$, a variant of $Pca2_1$ chosen for the consistency with the axes settings of *Pbca*), and the thermodynamically favored monoclinic (M: $P2_1/c$) phases[7,8]. Typically, $HfO_2$ thin films fabricated by ALD necessitate rapid thermal annealing (RTA) to stabilize the FE O-phase while suppressing T- and M-phases[6]. As a result, stabilizing the metastable FE O-phase poses a multifaceted challenge influenced by factors such as elemental doping[9-11], oxygen deficiency[12,13], mechanical strain[14,15], RTA temperature and cooling rate[16,17]. Additionally, variations in the metastable FE phase during working, specifically electric cycling, can lead to significant stability and reliability issues in devices. Notably, the domain structure of fluorite oxides in FE phase differs from that observed in conventional perovskite ferroelectrics. It is reported to possess a flat polar phonon band resulting in 180° domain with half-unit cell width[18,19]. Additionally, 90° ferroelastic domains and switching phenomena have also been observed[20,21]. The evolution of phase and domain structures under electric fields will undoubtedly impact device performance significantly[22]. Therefore, elucidating the relationship between these structural characteristics and FE behavior is essential for enhancing device functionality.

In FE memory, the FE capacitor (CAP) demonstrates compatibility with prevalent dynamic random-access memory (DRAM) device architectures and is considered to hold substantial promise for applications in FE memory[23]. Concerning the properties of HfO$_2$-based FE CAP, during electric cycling, the remnant polarization (representing memory window) exhibits a gradual increase, leading to the wake-up effect, followed by a subsequent decrease referred to as fatigue[24]. Factors influencing the memory window encompass phase transitions[25,26], alterations in domain orientation[27,28], variations in oxygen vacancies[29-31], and so on. Regarding the structural contributions, several studies proposed that wake-up occurs due to the transition from T-to-O phase[25], while fatigue arises from O-to-M martensitic transition[29,32] and reversible transitions between O-*Pbc*2$_1$ and O-*Pbca* phases[19]. The 90° polarization switching after poling may also contribute to the wake-up effect[33]. However, due to the challenges in characterizing ALD-prepared ultra-thin polycrystalline HfO$_2$-based thin films with polyphase coexistence and structural similarities, thus far, experimental evidence regarding structural evolution path under electric fields remains limited. Most investigations rely on macroscopic experiments, *ex-situ* structural representations and theoretical hypotheses.

Here, an O-*Pmn*2$_1$-like transient state is experimentally verified. By applying an *in-situ* electrical bias to the Hf$_{0.5}$Zr$_{0.5}$O$_2$ (HZO) FE CAP in spherical aberration-corrected transmission electron microscope (Cs-TEM), we directly observed that through an O-*Pmn*2$_1$-like transient state, non-FE O-*Pbca* and M-like misarrangement transform into the FE O-*Pca*2$_1$ phase. Furthermore, the polar axis of the FE-phase reorients in alignment with the direction of electric field through ferroelastic switching, resulting in enhanced FE polarization. As cycling progresses, however, there is a gradual collapse of the polar axis, leading to degradation of FE polarization. Through elucidating the structure evolution path of FE HZO nanocrystals under the electric field, the understanding of structural transformation in fluorite oxide-based FE materials could be deepened, thus offering insights for designing HfO$_2$-based FE memory.

**Unity of phase structure under the 1$^{st}$ bias.** The endurance test results of our TiN/HZO/TiN FE CAP (Figure S1, with device structure) illustrate that the HZO thin film fabricated by ALD in this study has normal ferroelectric behaviors, exhibiting characteristic wake-up and fatigue behaviors. To investigate the structure transitions in polycrystalline HZO FE thin films under applied electric fields, we conducted an *in-situ* Cs-TEM biasing experiment, as illustrated in **Figure 1a**. We defined the x, y and z axes to describe the typical polarization directions of FE

O-phases, with polarization along the y-direction (out of plane) being preferred for FE CAP applications. **Figure 1b** shows the high-resolution transmission electron microscope (HRTEM) image of the initial atomic structure of HZO before biasing. Additional scanning transmission electron microscope (STEM) high-angle annular dark-field (HAADF) and annular bright-field (ABF) images are provided in the supplementary material (Figure S2). Based on the HRTEM image, the corresponding fast Fourier transform (FFT) and the oxygen ions arrangement in the ABF image, it turns out that the initial HZO is mainly composed of the FE $O_x$-*Pbc*$2_1$ phase with polarization oriented along x. Nevertheless, other compositions remain non-negligible. The red bar-like area indicates the presence of one unit-cell width of M-like misarrangement[21], as illustrated in the inset HAADF image. A comparative analysis between the M-like misarrangement and standard M-phase is provided in Figure S3. Moreover, the cyan bar-like area marks two unit-cell widths of non-FE O-*Pbca* phase, with the inset showing their FFT and the 1/4 diffraction spots. Additionally, a small $O_z$ region is identified at the lower section of **Figure 1b**. Therefore, the initial HZO is a mixed structure consisting of $O_x$, $O_z$, O-*Pbca*, and M-like misarrangement phases.

Subsequently, direct current (DC) bias perpendicular to the major polarization direction of the initial HZO FE CAP was applied. As depicted in **Figure 1c**, the $O_x$, $O_z$, O-*Pbca*, and M-like misarrangement collectively transformed into a transient state characterized by evenly spaced atomic planes with an angle of 90°, accompanied by the disappearance of FFT superlattice diffraction spots. When the voltage was reduced to 0 V (**Figure 1d**), the entire HZO nanocrystal evolved into a uniform $O_x$ phase. The simulated FFTs for various structures are shown in Figure S4, which assists in identifying individual phases. **Figure 1e** provides a schematic diagram summarizing the atomic structural transformations during biasing. The transition from non-FE phases (O-*Pbca* and M-like misarrangement) to FE phase ($O_x$-*Pbc*$2_1$) and the unification of polarization orientation from $O_z$-to-$O_x$ both enable a potential ferroelectricity enhancement in HZO CAP, which should be an essential primary step in initiating the wake-up process.

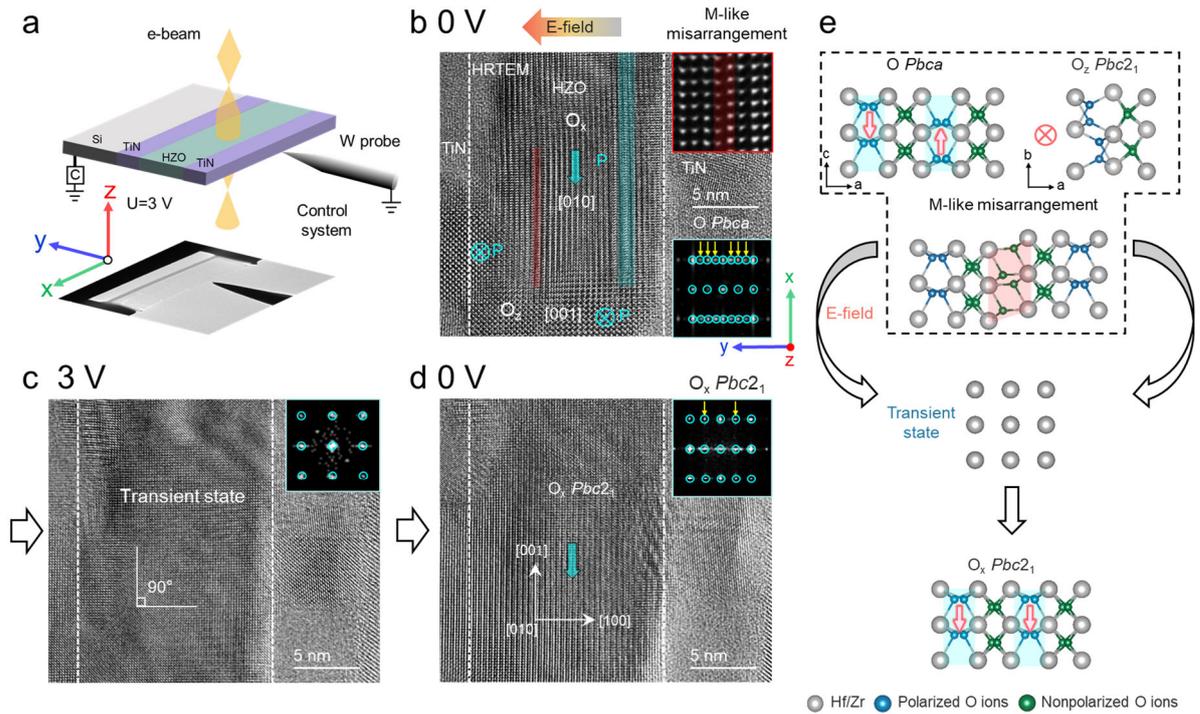

**Figure 1 | Unity of phase structure under the 1st bias. a,** Schematic diagram of the *in-situ* Cs-TEM biasing system for TiN/HZO/TiN capacitor. **b,** HRTEM image depicting the coexistence of $O_x/O_z$-$Pca2_1$ phase, O-*Pbca* phase and M-like misarrangement in the initial state. The O-*Pbca* is highlighted in cyan, and the M-like misarrangement is indicated in red. The cyan arrows denote the FE polarization direction. The orange arrow illustrates the electric field direction. The insets show the STEM-HAADF image of M-like misarrangement and the FFT analysis corresponding to O-*Pbca*. **c,** HRTEM image of the transient state appearance under a 3 V electric field, where the inset is the corresponding FFT image. The atomic planes present a 90° angle. **d,** HRTEM image of full $O_x$-$Pbc2_1$ after biasing. The inset shows the corresponding FFT image. Characteristic diffraction spots are denoted by yellow arrows. **e,** Schematics of the atomic structure evolution under 1st bias.

**Polarization enhancement under the 2nd bias. Figure 2a-2c** present the HRTEM images of the same HZO nanocrystal (as that in **Figure 1**) under 2nd bias, extracted from Supplementary Video 1, providing a visual representation of the dynamic structure evolution. In **Figure 2a**, a 3V DC electric field induces a transition from $O_x$ to the transient structure again. **Figure 2b-2c** illustrate subsequent structural transformations at time intervals of 0.1 s and 0.2 s, respectively. Analysis of the lattice and the characteristic 1/2 FFT diffraction spots reveals the formation of $O_x$ and $O_y$ domains with *Pbc*2$_1$ symmetry within 0.1 s. Notably, the $O_y$ domain aligned with the direction of the applied electric field, yielding a formation of 90° FE domain wall relative to $O_x$. During the transition at 0.2 s, both $O_x$ and $O_y$ FE domains gradually expanded, accompanied by the movement of this 90° domain wall. The observed reorientation phenomenon, where polarization changed towards alignment with the applied electric field, is particularly relevant as it plays an essential role in contributing to the wake-up effect commonly seen in FE CAP.

Nevertheless, the observed reorientation process is not instantaneous. As depicted in **Figure 2d**, the length ratios (a/b) between the polar sublattice (a, purple box) and nonpolar sublattice (b, pink box) reveal that the a/b at 0.1 s is smaller than that of the standard O-*Pbc*2$_1$, the a/b at 0.2 s has approached that of the standard O-*Pbc*2$_1$, suggesting a gradual transition from the transient state towards stable configurations. During this interval, another significant finding emerges at the atomic scale that affects domain dynamics. The lattice framework as illustrated in **Figure 2d** (indicated by red borders) for the transient state, $O_x$ and $O_y$ at time intervals of 0.1 s and 0.2 s respectively, remains nearly unchanged throughout the entire transition process, likely due to a constrained strain environment limiting crystal structure mobility. Such constraints may hinder rapid atomic rearrangements essential for structure transformation. Therefore, although reorientation contributes to polarization, the wake-up process of FE CAP typically requires an extended time duration of ~$10^4$ electric cycles, which is believed to be associated with this constrained lattice framework. $O_x$ is more consistent with the standard O-*Pbc*2$_1$, while $O_y$ forms a co-lattice domain wall alongside $O_x$ and is influenced by the lattice constraint. Notably, the lattice constants of the transient state, $O_x$ and $O_y$ in the x-direction are consistently larger than those in the y-direction, indicating that the HZO nanocrystal tends to remain in its initial $O_x$ state rather than transitioning into a constrained $O_y$ phase. Therefore, it follows that the contribution from the reoriented FE-phase towards polarization is in general temporary, and is subject to degradation with continued electrical cycles.

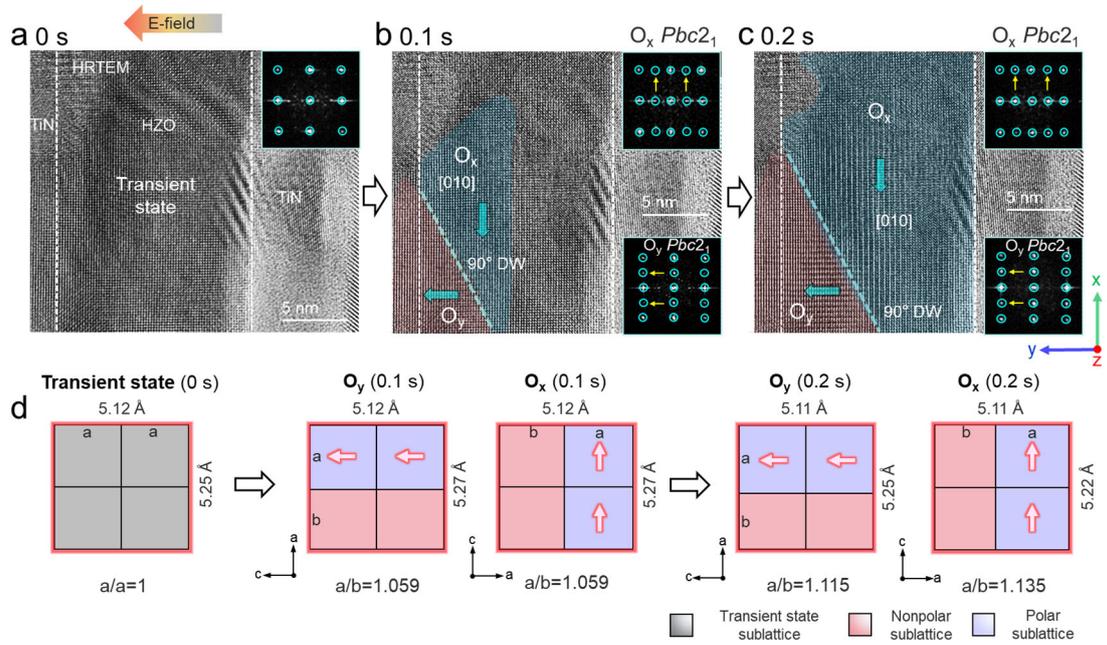

**Figure 2 | Polarization enhancement under the 2nd bias. a,** HRTEM image of the transient state under the 2nd 3 V electrical bias. **b-c,** Time-lapse HRTEM images of O-FE domains at 0.1 s and 0.2 s, respectively, in which, $O_x$-$Pbc2_1$ and $O_y$-$Pbc2_1$ are indicated in blue and red, respectively. The 90° domain walls are denoted by cyan dashed lines. The cyan arrow represents the polarization direction. The insets show the corresponding FFT patterns. **d,** Schematic of the lattice structure evolution from transient state (0 s) to $O_x$/$O_y$ formation (0.1 s), followed by $O_x$/$O_y$ expansion (0.2 s). The lattice frameworks are outlined with red borders, with the lattice parameters clearly marked. The ratios between the lengths of nonpolar sublattice a and polar sublattice b are also indicated.

**Polarization degradation under the 3rd and 4th electrical bias.** The same HZO nanocrystal was subjected to continuous *in-situ* bias stimulation to explore the further evolutionary pathways of its atomic structure. Under the 3rd 3 V DC bias (**Figure 3a**), the $O_x$ and $O_y$ domains underwent a synergistic transformation through an identical transient state (Figure S5a) as previously observed, and subsequently reverting to a single $O_x$ domain. The disappearance of $O_y$ structure signifies a degradation in FE polarization. During the subsequent 4th bias, the $O_x$ phase again underwent the identical transient state (Figure S5b) into an $O_z$ lattice configuration. The three distinct $O_z$ regions are labeled as 1, 2, and 3, with their corresponding FFT patterns providing crystallographic insights. **Figure 3c** shows the STEM-HAADF image extracted from the white border in **Figure 3b**, with cyan dashed lines denoting the domain walls formed by distinct $O_z$ domains. The atomic structure analysis reveals mirror symmetry between regions 1 and 2, alongside a notable 90° rotational variance for region 3. This unusual structural evolution leads to the collapse of O-phase polar axis, thereby diminishing its contribution to FE polarization. The emergence of $O_z$ phase within HZO film exhibits parallels to the polarization degradation mechanism in perovskite FEs, as reported by Huang et al., which is also attributed to the formation of $O_z$ domains under multiple electric field cycles[34]. Therefore, this work provides robust experimental evidence for understanding the fatigue mechanism in fluorite-type FE oxide materials.

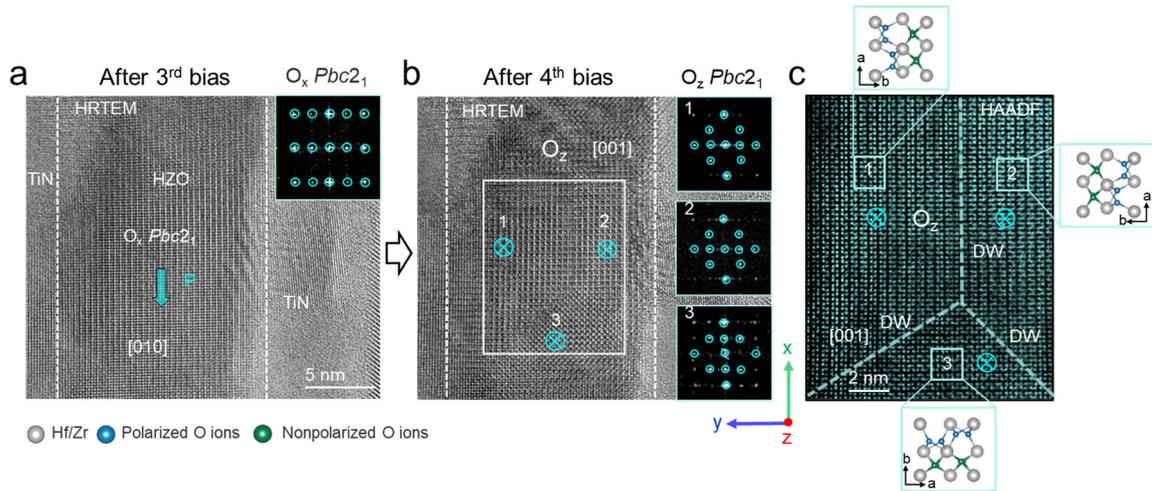

**Figure 3 | Polarization degradation under the 3$^{rd}$ and 4$^{th}$ electrical bias. a,** HRTEM image of the $O_x$-$Pbc2_1$ formation after the 3$^{rd}$ bias. **b,** HRTEM image shows the coexistence of three $O_z$-regions in response to the 4$^{th}$ electric field stimulation. The insets show the corresponding FFT patterns with characteristic diffraction spots highlighted. **c,** STEM-HAADF image shows atomic structures of $O_z$ 1 to 3, obtained from the white border in panel b. The domain walls are denoted by cyan dashed lines. Cyan arrows represent polarization directions.

**Theoretical explanation of the transient state.** Throughout each biasing process, we consistently observed the emergence of a transition state characterized by equidistant atomic planes, exemplified by the reversible transition between $O_x$-$Pbc2_1$ and $O_y$-$Pbc2_1$ via this transition state (**Figure 4a**). However, the crystal structure of this transition state remains elusive due to the inherent limitations of HRTEM in accurately depicting the configuration of oxygen ions within the lattice, as well as the fleeting nature of the transition state itself. Therefore, we conducted comprehensive first-principles calculations aiming at elucidating both the transient state structure and its formation under electric field stimulation. We employed both variable cell nudged elastic band (VC-NEB) and climbing image nudged elastic band (CI-NEB) methods[35-37] to compute the energy variations along the transition pathway from $O_x$-$Pbc2_1$ to $O_y$-$Pbc2_1$, including identifying intermediate transient states and their associated energy barriers. During VC-NEB calculations, both atomic positions and lattice parameters are adjusted to find the global energy minimum. **Figure 4b** shows the energy diagram (red lines) obtained from the VC-NEB calculation. There is an energy saddle point, and the corresponding intermediate transition state structure has been extracted. Notably, this transition state matches the structure of the non-centrosymmetric O-$Pmn2_1$ phase, which has also been proposed as a candidate in the search for low-energy FE phases using density function theory (DFT)[8,38,39]. However, to date, the role of O-$Pmn2_1$ phase in hafnia-based FEs has not been experimentally confirmed. The polarization vector (blue arrows) of the O-$Pmn2_1$ transition state is aligned along the diagonal of the lattice (45°), with a polarization magnitude of 0.56 C/m$^2$. However, the angle between the two atomic planes within the O-$Pmn2_1$ structure measures 84.62°, which deviates from the observed value of 90° in this experiment (**Figure 1c**). Given that strain confines the lattice framework, employing CI-NEB method with a relatively fixed lattice is a more effective approach for calculating such transitions. **Figure 4c** presents the energy diagram (blue lines) obtained from the CI-NEB calculation. The transient state exhibits a polar O-$Pmn2_1$-like (distorted O-$Pmn2_1$) structure. The displacement of the polarized oxygen ion decreases, and the angle between the atomic planes becomes 90°. After structure relaxation by first-principles, the O-$Pmn2_1$-like structure transforms into the standard $Pmn2_1$ with the same symmetry. Tables S1, S2 in supplementary material provide the lattice constants and atomic positions of O-$Pbc2_1$, O-$Pmn2_1$, respectively. Therefore, the FE domain reorientation under electric field does not occur directly, but through a 45° polarized O-$Pmn2_1$-like transient phase on the path of transition from $O_x$ to $O_y$ structures, before undergoing a subsequent 90° polarization transition to $O_y$ as illustrated in **Figure 4d**.

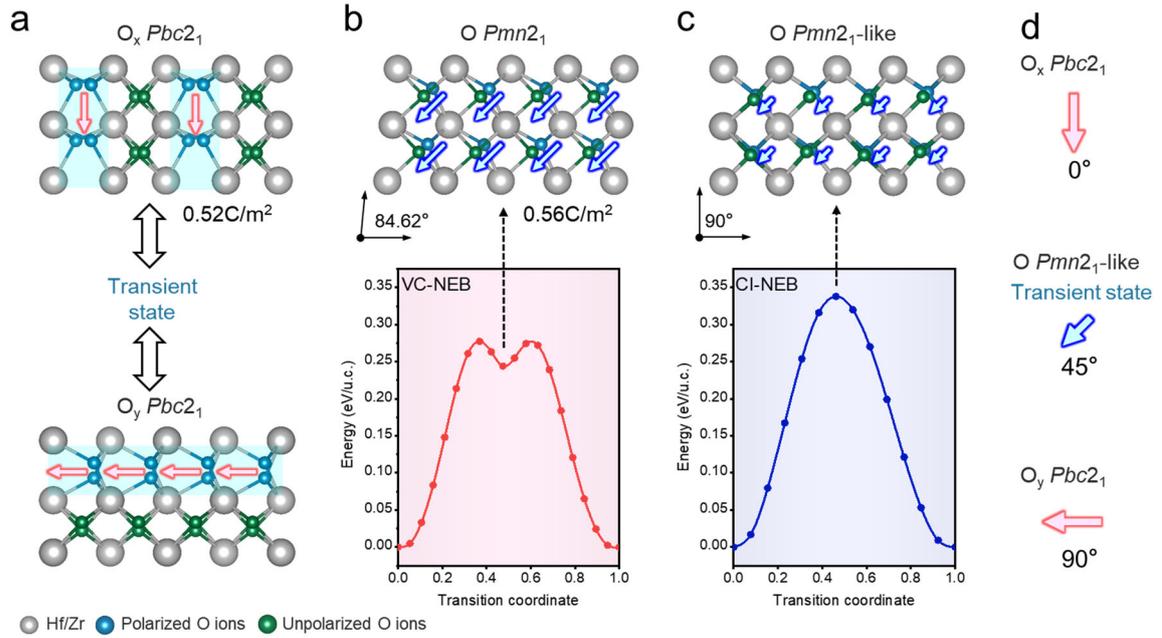

**Figure 4 | Theoretical explanation of the transient state. a,** Atomic models of the structure transition from $O_x$-$Pbc2_1$ to $O_y$-$Pbc2_1$ through a transient state. The polarization vectors are indicated by red arrows. **b-c,** Atomic structures of O-$Pmn2_1$ and O-$Pmn2_1$-like transient states extracted from the $O_x$-to-$O_y$ transition energy diagram, which are calculated by VC-NEB (red lines) and CI-NEB (blue lines), respectively. The polarization vectors for polar transient states are indicated by blue arrows. **d,** Schematic of polarization vectors evolution during the above structure transition.

In conclusion, we conducted a comprehensive investigation into the evolution path of the atomic-scale structure of HZO nanocrystals by employing *in-situ* techniques to apply an electric field to FE CAP. Our observations revealed that the transition from nonpolar to polar phases, along with the reorientation of FE domains, involves complex atomic-scale mechanisms that govern the FE CAP properties. Among them, the polar axis changes towards alignment with electric field contributes to the wake-up effect observed in HZO FE CAP. Furthermore, the emergence of the $O_z$ phase plays a significant role in device fatigue, and the fatigue behavior is also related to the constrained lattice framework. Importantly, an O-$Pmn2_1$-like transient state structure is experimentally confirmed to exist during each phase and domain evolution under electric field, which exhibits unique characteristic that is crucial for understanding the domain dynamics. A 45° polarization associated with this O-$Pmn2_1$-like transient state structure is supposed to facilitate polarization switching and modify polarization directions. This study uncovers not only the structural evolution path under electric field, but also their implications of structural switching on FE properties for ALD-prepared $HfO_2$-based thin films, which are of great importance in guiding process optimization and performance improvement in $HfO_2$-based FE community.

**Methods**

**Sample preparation:** A 10 nm TiN bottom electrode was deposited on silicon (Si) substrate via ion beam sputtering. For the deposition of a 15 nm thick HZO film, $Zr[N(C_2H_5)CH_3]_4$ and $Hf[N(C_2H_5)CH_3]_4$ were selected as the atomic layer deposition (ALD) precursors. Following the HZO deposition, a 40 nm thick TiN film was deposited as the top electrode. To achieve the FE properties, rapid thermal annealing (RTA) at 550 °C for 30 s in an $N_2$ ambient environment was conducted. For the *in-situ* spherical aberration-corrected transmission electron microscope (Cs-TEM) experiments, the TiN/HZO/TiN nanosheet capacitor unit was fabricated on a copper (Cu) grid using a Thermo Fisher Scientific Helios G4 focus ion beam (FIB) instrument.

***In-situ* Cs-TEM experiments:** The high-resolution transmission electron microscope (HRTEM) and scanning transmission electron microscope (STEM) analyses were performed using JEOL JEM-ARM300F microscope with double Cs-correctors, operating at 300 kV. The TiN/HZO/TiN nanosheet capacitor was placed on the front end of X-nano TEM holder. In the *in-situ* biasing experiment, a tungsten (W) tip was maneuvered in three-dimension (3D) to establish contact with the nanosheet capacitor, thereby forming an electrical circuit among the capacitor, the W tip and the external electrical facility (Keithley 2450).

**First-principles calculations:** Density functional theory (DFT) calculations were performed within the Vienna ab initio simulation package (VASP) using the projector-augmented wave (PAW) method[40,41]. The valence electron configurations were 2*s* and 2*p* for oxygen; 5*p*, 5*d*, and 6*s* for hafnium. The Perdew-Burke-Ernzerhof functional of the generalized gradient-approximation type (GGA-PBE) was adopted to treat the exchange-correlation. A cutoff energy of 600 eV and a Monkhorst Pack k-mesh were used in the calculations. The atomic positions and lattice constants were fully relaxed until the atomic Feynman forces were smaller than 0.01 eV/Å in each direction. Both the variable-cell nudged elastic band (VC-NEB) and conventional climbing image nudged elastic band (CI-NEB) methods are employed to simulate the experimentally observed phase transition process. The minimum energy paths for phase transition were obtained until the atomic Feynman forces were smaller than 0.02 eV/Å in all directions, and the energy convergence was below a $10^{-5}$ eV energy difference between two consecutive self-consistent steps. The Berry phase method was used to calculate the spontaneous polarization[42].

**Supplementary Information**

Supplementary information is available from the Online content.

**Declaration of competing interest**

The authors declare no competing financial interests.

**Acknowledgment**

This work is supported by the National Natural Science Foundation of China (grant nos. 62174054, 92064003, 62104071 and 12134003), the Natural Science Foundation of Shanghai (23ZR1418000), the Shanghai Science and Technology Innovation Action Plan Intergovernmental International Science and Technology Cooperation Program (23520712000), the Strategic Priority Research Program of the Chinese Academy of Sciences (grant no. XDB44000000), Open Research Fund of China State Key Laboratory of Materials for Integrated Circuits(No. NKLJC-K2023-05) and China Postdoctoral Science Foundation (2024M750924).

# Supplementary Information for

Structure evolution path of ferroelectric hafnium zirconium oxide nanocrystals under *in-situ* biasing

Yunzhe Zheng[1†], Heng Yu[2†], Tianjiao Xin[1,3†], Kan-Hao Xue[2]*, Yilin Xu[1], Zhaomeng Gao[1], Cheng Liu[1], Qiwendong Zhao[1], Yonghui Zheng[1], Xiangshui Miao[2], Yan Cheng[1,3]*

[1]Key Laboratory of Polar Materials and Devices (MOE), Department of Electronics, East China Normal University, Shanghai 200241, China.

[2]School of Integrated Circuits, Huazhong University of Science and Technology, Wuhan 430074, China.

[3]State Key Laboratory of Materials for Integrated Circuits, Shanghai Institute of Microsystem and Information Technology, Chinese Academy of Sciences, Shanghai 200050, China.

[†]Yunzhe Zheng, Heng Yu, and Tianjiao Xin contributed equally.

*Corresponding author: ycheng@ee.ecnu.edu.cn; xkh@hust.edu.cn.

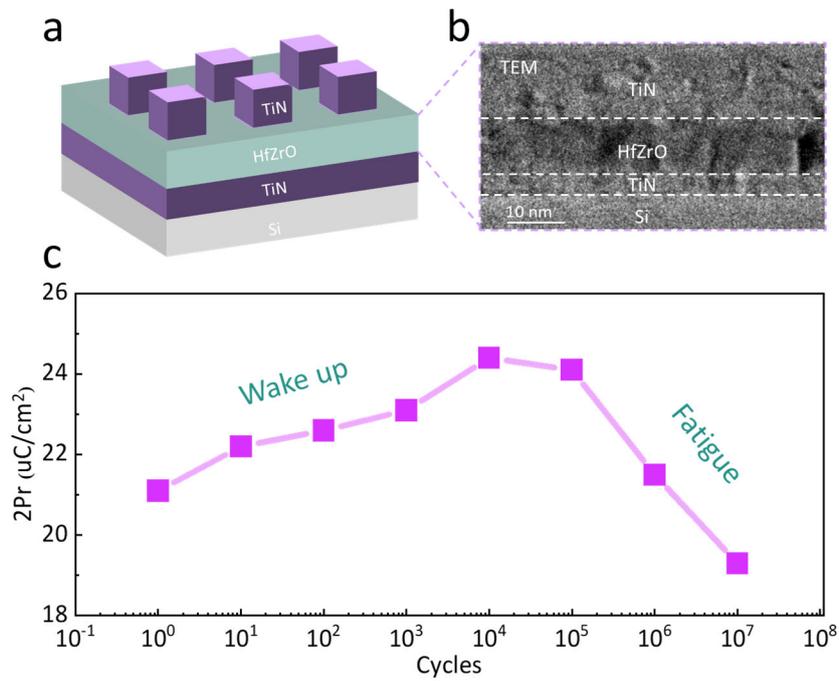

**Figure S1 | a,** Schematic diagram of TiN/Hf$_{0.5}$Zr$_{0.5}$O$_2$ (HZO)/TiN ferroelectric (FE) capacitor (CAP) on silicon (Si) substrate. **b,** Transmission electron microscopy (TEM) images of TiN/HZO/TiN FE CAP. **c,** 2Pr of HZO CAP with electrical cycling, the HZO thin film prepared by atomic layer deposition (ALD) in this study has good ferroelectricity, and shows typical polarization enhancement (wake-up) and degradation (fatigue) phenomena.

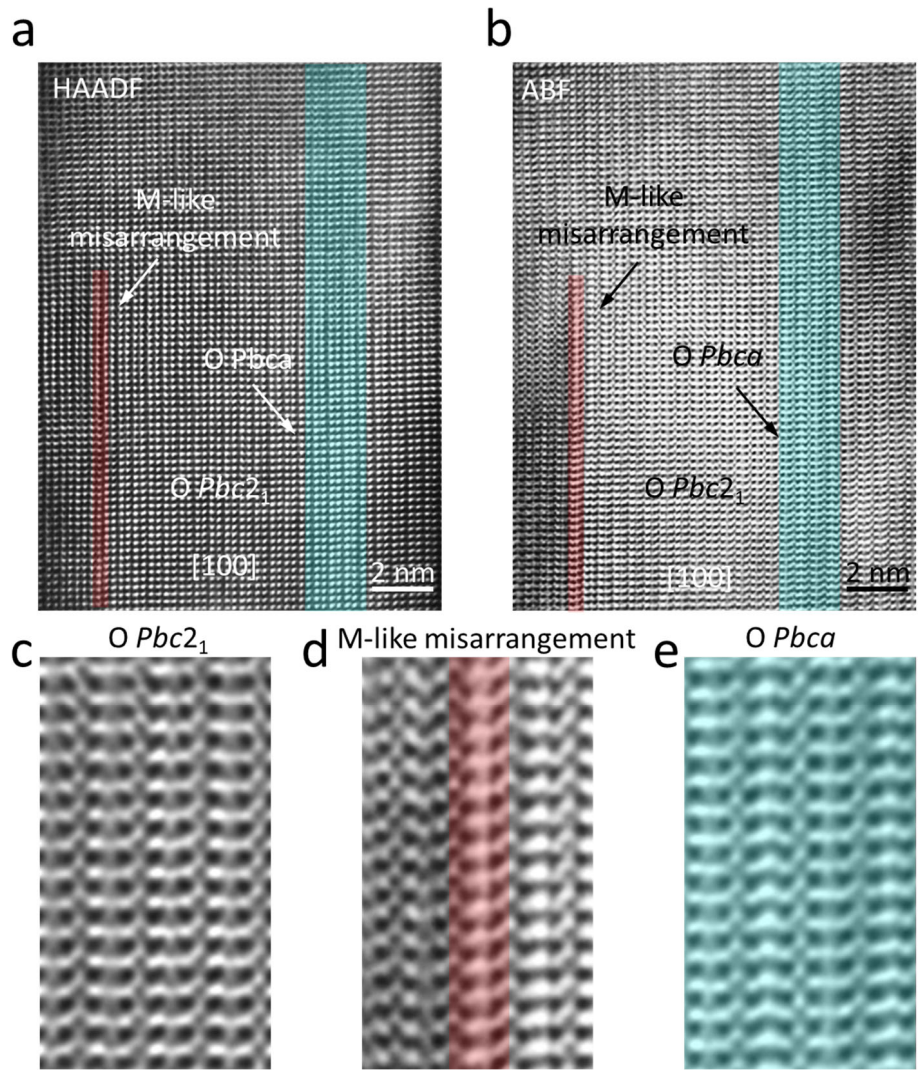

**Figure S2 | a, b,** The STEM high-angle annular dark-field (HAADF) and annular bright-field (ABF) images corresponding to HZO nanocrystals in the pristine state as shown in Figure 1b. The O-*Pbca* phase is marked cyan, and the M-like misarrangement phase is marked red. **c,** The ABF images of the O-*Pbc*2$_1$ phase, **d,** the M-like misarrangement and **e,** the O-*Pbca* phase, extracted from (**b**).

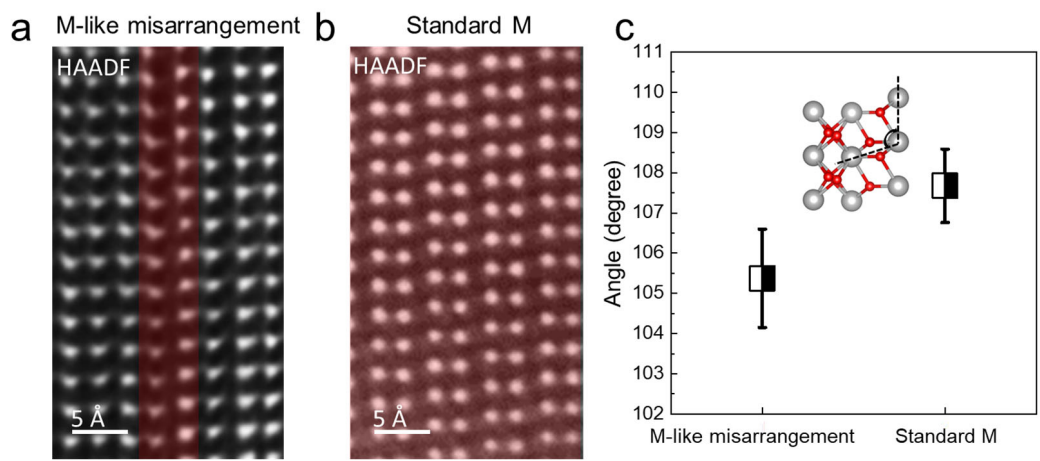

**Figure S3 | a,** The HAADF image of the M-like misarrangement. **b,** The HAADF image of the standard M-phase. **c,** The comparison between the M-like misarrangement and the standard M-phase. We measured the angle between the atomic lines shown in the inset. The angle measured for the M-like misarrangement was around 106.38°, while the angle for the standard M-phase was around 108.68°. According to recent research, the distorted M-phase is more easily transformed into the FE O-phase than the standard M-phase[1].

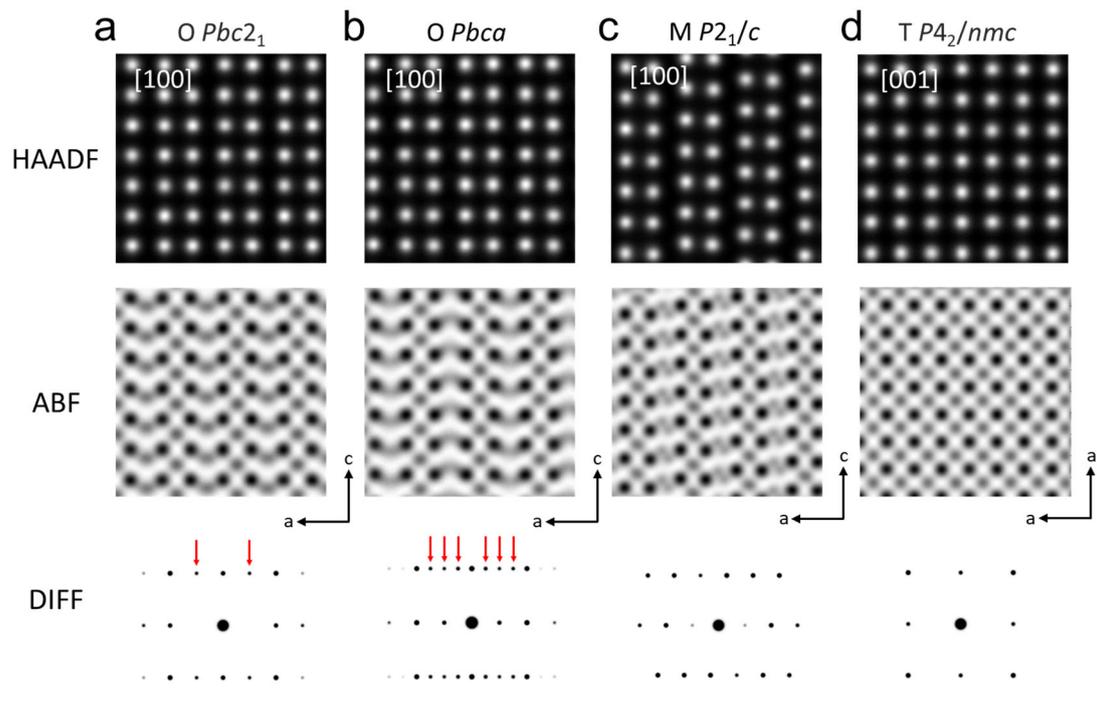

**Figure S4** | Simulated HAADF, ABF images and diffraction patterns of (**a**) O-*Pbc*2$_1$, (**b**) O-*Pbca*, (**c**) O-*P*2$_1$/*c* and (**d**) T-*P*4$_2$/*nmc* phase. The STEM images were generated using commercial software (MacTempas) based on the multislice method.

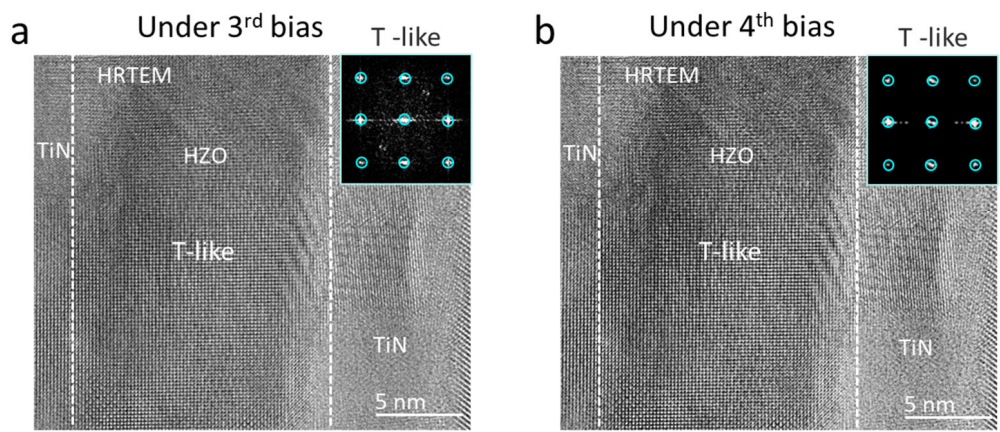

**Figure S5** | High-resolution transmission electron microscope (HRTEM) image of the T-like transient structure appearance under the (**a**) 3rd and (**b**) 4th electric bias, respectively. The insets show the corresponding FFT images with the characteristic diffraction spots.

**Table S1.** Atomic positions of *Pbc*2₁ structure, with lattice constants a = 5.268 Å, b = 5.048 Å, c = 5.078 Å, α = 90.000°, β = 90.000°, γ = 90.000°.

|     | x     | y     | z     |
| --- | ----- | ----- | ----- |
| a   | 5.268 | 0.000 | 0.000 |
| b   | 0.000 | 5.048 | 0.000 |
| c   | 0.000 | 0.000 | 5.078 |
| Hf  | 0.434 | 0.000 | 0.500 |
| Hf  | 0.500 | 0.467 | 0.000 |
| Hf  | 0.000 | 0.000 | 0.000 |
| Hf  | 0.934 | 0.467 | 0.500 |
| O   | 0.197 | 0.271 | 0.750 |
| O   | 0.737 | 0.196 | 0.250 |
| O   | 0.697 | 0.196 | 0.750 |
| O   | 0.237 | 0.271 | 0.250 |
| O   | 0.332 | 0.801 | 0.857 |
| O   | 0.102 | 0.801 | 0.357 |
| O   | 0.832 | 0.665 | 0.857 |
| O   | 0.602 | 0.665 | 0.357 |

\

**Table S2.** Atomic positions of *Pmn*2$_1$ structure, with lattice constants a = 5.180 Å, b = 5.119 Å, c = 5.119 Å, α = 84.345°, β = 90.000°, γ = 90.000°.

|     | x     | y     | z     |
| --- | ----- | ----- | ----- |
| a   | 5.180 | 0.000 | 0.000 |
| b   | 0.000 | 5.119 | 0.000 |
| c   | 0.000 | 0.504 | 5.094 |
| Hf  | 0.476 | 0.000 | 0.500 |
| Hf  | 0.476 | 0.500 | 0.000 |
| Hf  | 0.000 | 0.000 | 0.000 |
| Hf  | 0.000 | 0.500 | 0.500 |
| O   | 0.294 | 0.253 | 0.753 |
| O   | 0.669 | 0.327 | 0.326 |
| O   | 0.807 | 0.327 | 0.826 |
| O   | 0.183 | 0.253 | 0.253 |
| O   | 0.183 | 0.753 | 0.753 |
| O   | 0.294 | 0.753 | 0.253 |
| O   | 0.669 | 0.827 | 0.827 |
| O   | 0.807 | 0.827 | 0.327 |